\documentclass[]{interact}

\usepackage{epstopdf}
\usepackage{subfigure}
\usepackage{amssymb}
\usepackage{graphicx}
\usepackage{color}
\usepackage{amsmath}

\usepackage{natbib}
\bibpunct[, ]{(}{)}{;}{a}{}{,}
\renewcommand\bibfont{\fontsize{10}{12}\selectfont}

\theoremstyle{plain}

\theoremstyle{definition}

\theoremstyle{remark}

\begin{document}
\articletype{ARTICLE TEMPLATE}
\title{ISAR imaging of space objects using encoded apertures}

\author{
\name{M. Roueinfar\textsuperscript{a} and M.H. Kahaei\textsuperscript{b}\thanks{CONTACT M.H. Kahaei. Email: kahaei@iust.ac.ir}}
\affil{\textsuperscript{a,b}School of Electrical Engineering, Iran University of Science and Technology,Tehran,Iran}
}
\maketitle

\begin{abstract}
A major threat to satellites is space debris with their low mass and high rotational speed. Accordingly, the short observation time of these objects is a major limitation in space research for appropriate detection and decision. As a result, these objects not fully illuminated, leading to their incomplete images at any snapshot. In this paper, we propose a method to decrease the number of snapshots in a given observation time and using a limited number of spot beams per snapshot called the encoded aperture.  To recover the space debris images,  an inverse problem is defined based on compressive sensing methods. Also, we show that for satellite imaging the $TV$ norm is more appropriate. We develop a procedure to recover space debris and satellites using $L_{1}$ and $TV$ norms. Using simulation results, we compare the results with the well-known SBL and $SL_{0}$ norm in terms of the number of snapshots, MSE, SNR, and running time. It is shown that our proposed method can successfully recover the space objects images using a fewer number of snapshots.
\end{abstract}
\section{Introduction}
ISAR is an imaging radar for identifying, recognizing, and distinguishing moving objects from each other `\citep{Chn14}'. This type of radar can extract a 2D image of  space objects in both range and azimuth (or cross-range) directions. In addition, it can work in almost all-weather conditions, which makes it an appropriate choice for imaging of rotating objects in some problems `\citep{Ozd14}' such as space debris imaging. These objects may lie in different orbits due to the remaining parts of disabled satellites, spacecrafts, meteors, and etc. The dimensions  of space debris are very diverse and due to their low mass, they can move at high speeds.
As a result, an important challenge for spacecraft or satellite launching is their impact on orbital debris which could create expensive economical and research costs `\citep{Ang19}'. Accordingly,  space debris imaging has received increasing attention in space research for  which several optical or radar-based methods have been reported `\citep{Sci17,Zhu15}'. In this regard, ISAR is a good choice for creating spatially images of space debris, which may also be used for imaging and monitoring active satellites.
To make acceptable images of space debris, two major issues are important to note. First, these objects have a high rotational speed due to their low mass, which leads to generating high-bandwidth Doppler frequency. Secondly,  the low observation time of space objects in orbit is a critical limitation in ISAR imaging and thus an important motivation for reducing the processing time  `\citep{Zhu15}'.
In order to increase the resolution of ISAR images, given the sparse nature of space debris images, Compressive Sensing (CS) methods are reasonable techniques to use. In `\citep{Hua14}', a method based on randomly stepped frequency radar is presented by benefiting from sparse recovery and CS, which improves the range and azimuth resolutions.  Also, a method is addressed in `\citep{kan21}' based on orthogonal coding signals with different delays and a modified Smoothed $L_{0}$-norm (SL0) algorithm.  By combining the motion characteristics of space debris, a method  is presented in `\citep{Zhu15}' for extracting high-resolution images using CS techniques. The method of Multiple Azimuth Beams (MAB) addressed in `\citep{Che18}' allows ISAR imaging of a large area using  simple processing, electronic scan, and digital beamforming. Also, `\citep{Wan18}' has recovered the scatterers using consecutive observations of space debris based on CS.  Moreover, radar imaging for spinning space debris with a dimension smaller than the radar range resolution is developed in `\citep{Wan10}' and `\citep{Bas19}'.
 The problem we are focusing on is the rapid rotation of space debris, which requires having enough snapshots to fully recover the image. On the other hand, due to the low observation time of space debris, access to a sufficient number of snapshots in a limited time is problematic. To solve this problem and improve the quality of the retrieved images, we use several snapshots. Also,  we evaluate the performance of the $L_{1}$ and $TV$ norms to more successfully recover ISAR images of satellite and space debris.
The paper is organized as follows. In Section 2, we introduce the signal model and ISAR imaging method based on active spot beams. Section 3 describes how randomly encoded apertures are generated in each snapshot. In Section 4, we use $L_{1}$ and $TV$ norms to retrieve ISAR images. Simulation results are given in Section 5, and Section 6 concludes the paper.
\section{Signal Model }
According to `\citep{Zhu15}', the baseband echo or the received signal after de-chirping and translational motion compensation consists of $K $ scatterers as 
\begin{equation}
\label{eq:e1}
S_{r}(t_{m})=\sum_{k=1}^{K-1}
\sigma_{k}\exp[j\frac{4\pi r_{k}\cos(\omega{t_{m}+\phi_{k}})}{\lambda}],
\end{equation}
where $\lambda$ is the radar signal wavelength,  $\sigma_{k}$  is the reflection coefficient of the $\textit{k}$th scatterer, $r_{k}$ and $\varphi_{k}$  show the respective polar coordinates, $\omega$ is the angular velocity, and $t_{m}$ denotes the slow time, $i.e.$, the time of received signal in azimuth direction. Then, the instantaneous phase of the $\textit{k}$th scatterer is given by
\begin{equation}
\label{eq:e2}
\theta(t_{m})=\frac{4\pi r_{k}\cos(\omega{t_{m}+\phi_{k}})}{\lambda},
\end{equation}
where the instantaneous phase derivative is defined as
\begin{equation}
\label{eq:e3}
\theta_{d}(t_{m})=\frac{d\theta(t_{m})}{dt_{m}}.
\end{equation}
Next, given that $ \theta_{d}(t_{m}) = 2 \pi f_{d} $, we get
\begin{equation}
\label{eq:e4}
f_{d} = \frac{d\theta(t_{m})}{2 \pi dt_{m}} = \frac{-2 r_{k}\omega\sin(\omega{t_{m}+\phi_{k}})}{\lambda},
\end{equation}
where the Doppler bandwidth of the received signal is `\citep{Zhu15}'
\begin{equation}
\label{eq:e5}
\Delta f_{d} =\frac{4 r_{k}\omega}{\lambda}.
\end{equation}
As seen, large angular velocities produce large Doppler bandwidths `\citep{Zhu15}'
\section{ Proposed ISAR imaging method}
Due to the fast movement of space debris, its imaging  in a short observation time is difficult. Here, we present a method for imaging of satellite and space debris as follows:
\begin{itemize}
\item Instead of  a single snapshot for imaging, we use multiple snapshots.
\item To reduce the amount of data, computations, and processing time in each snapshot, we use randomly selected spot beams in each snapshot, which we call encoded aperture imaging.
\item We use the Bernoulli distribution to determine active/inactive spot beams in each encoded aperture.
\item We use $L_{1}$ and $TV$ norms for image recovery using a limited number of snapshots and spot beams, and compare their performance.
\end{itemize}

\subsection{Encoded Aperture}
To capture images of low-mass, high-speed, space-spinning debris, a single snapshot may not produce a high-quality image. This difficulty can be effectively compensated by using more snapshots, each of which with N spot beams radiating simultaneously. This, as a result, leads to needing more processing time for image recovery which might be undesirable during the orbit observation time. But the time interval for receiving echo from radar targets is determined by the Pulse Repetition Interval (PRI), which accordingly limits the
image processing time. As a solution, if we increase the PRI, the ambiguity in the Doppler measurement increases, which leads to losing the image quality[6]. To overcome this difficulty, in the proposed algorithm we generate a few randomly-produced spot beams in each snapshot using the Bernoulli distribution [24]. We name this type of snapshot "encoded-aperture snapshots" (or simply encoded apertures) to produce an incomplete image of an object. Then, by using a few encoded apertures in an inverse problem and some recovery methods, a complete and improved image is retrieved. To illustrate, Fig. 1 shows an encoded aperture consisting of three spot beams, and Fig. 2 illustrates M encoded apertures with five spot beams at each snapshot. The radar Half Power Beam Widths (HPBW) for each spot beam are defined in elevation and azimuth directions by  $\theta$ and $\varphi$, respectively.

\begin{figure}
\centering
\includegraphics[angle=0,width=0.4\textwidth]{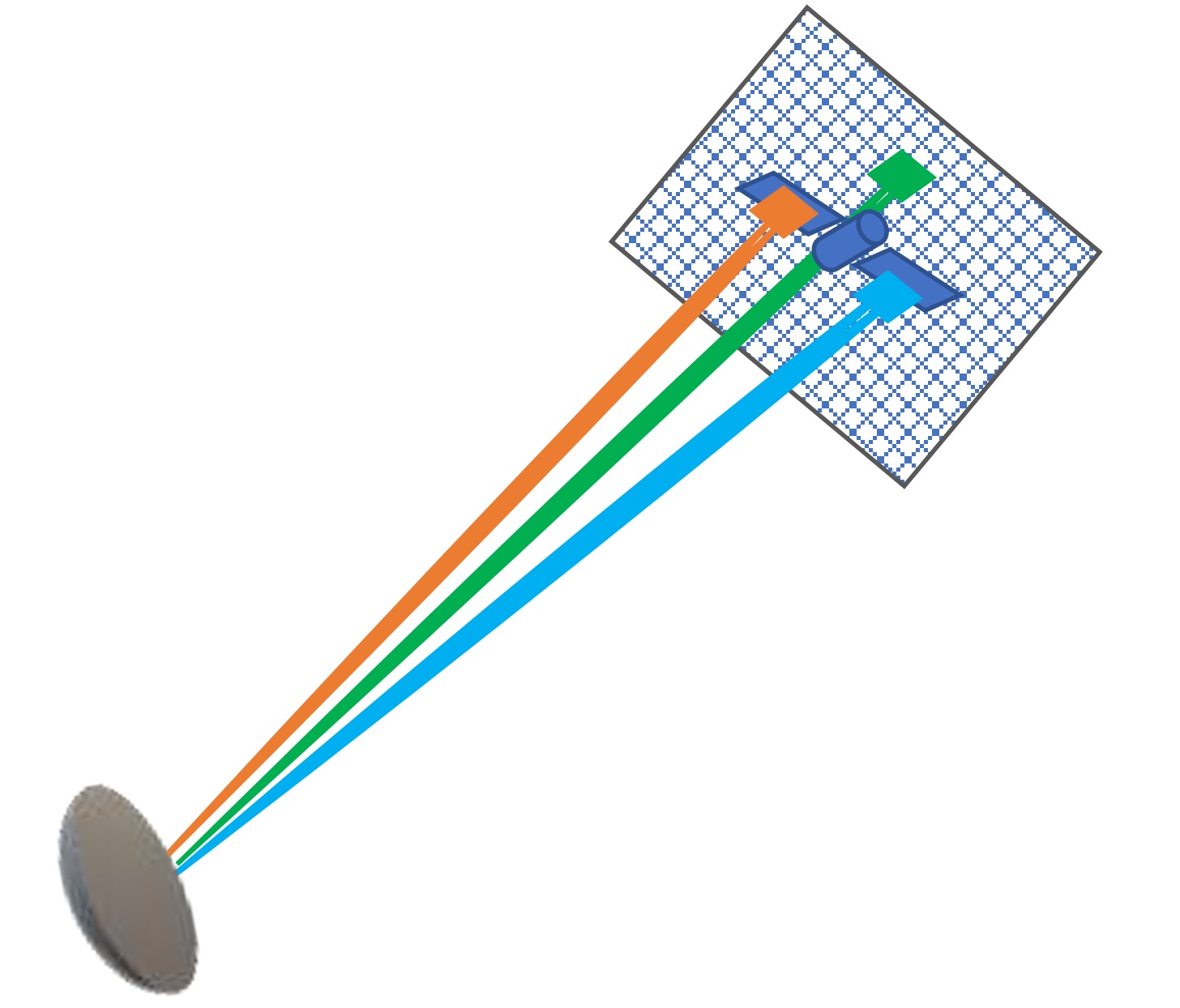}
\caption{ISAR imaging using a limited number of spot beams in each snapshot.}
\label{fig:Fig.ISAR}
\end{figure}
\begin{figure}
\centering
\includegraphics[angle=0,width=0.4\textwidth]{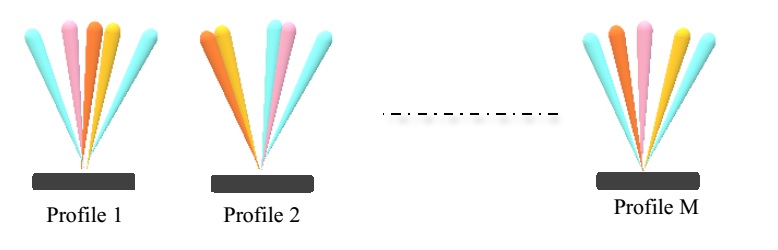}
\caption{Five different spot beams in $M$ snapshots.}
\label{fig:Fig.Spot}
\end{figure}
\subsection{Retrieval of ISAR images using CS}
\begin{figure}[t]
\centering
\includegraphics[angle=0,width=0.45\textwidth]{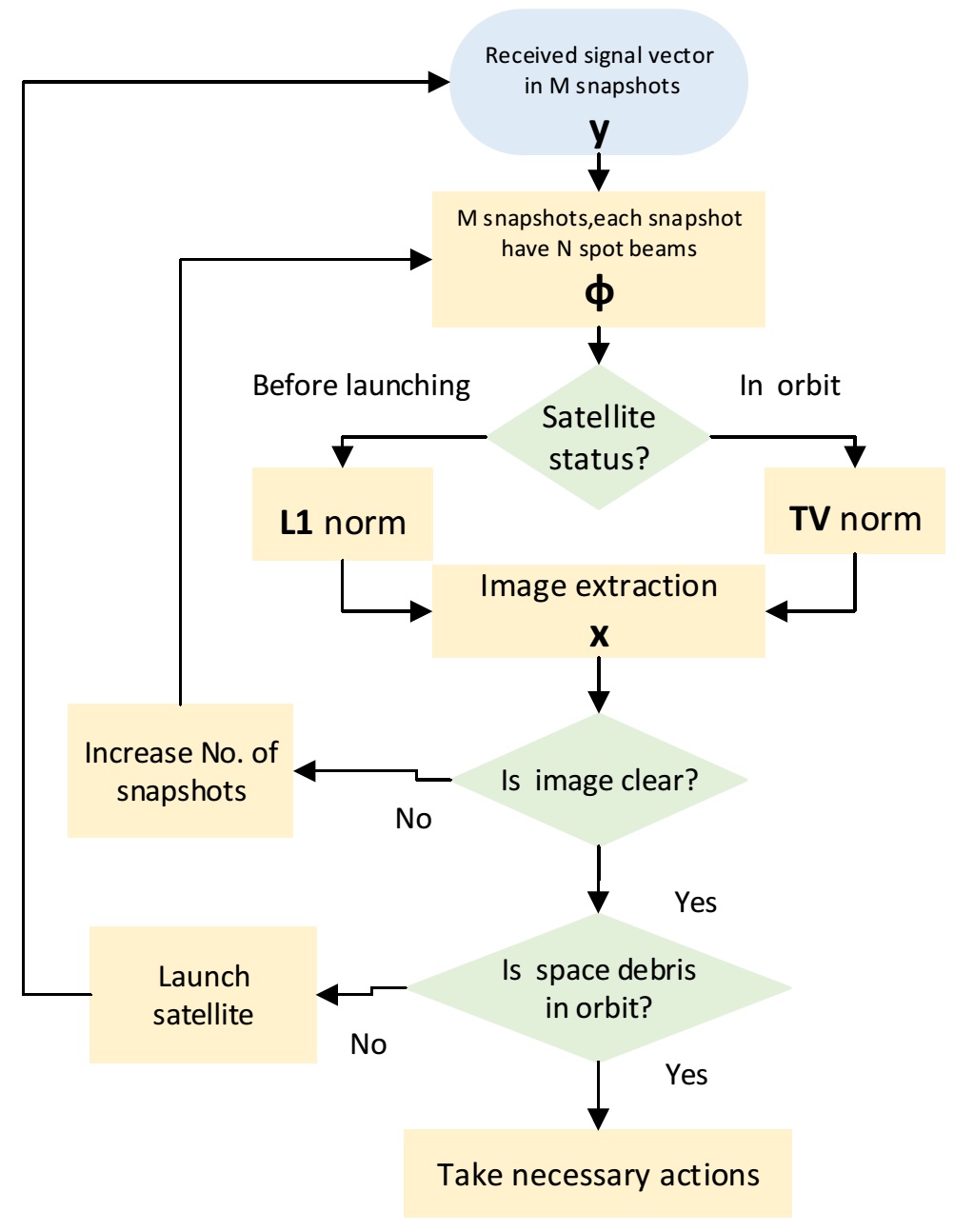}
\caption{Flowchart of applying the proposed method.}
\label{fig:Fig.Flow}
\end{figure}

We assume a space object image $ {\boldsymbol X}  \in  C ^ {n\times n} $ is observed at each encoded aperture by the spot beams in both range and azimuth directions  with $n$ showing the maximum number of spot beams in each direction. We vectorize ${\boldsymbol X}$  as ${\boldsymbol x}  \in  C ^ {N} $ with $r$ non-zero elements showing the sparsity of this vector and $N=n\times n$. Here, we use  $M$ encoded aperture, each of which  is randomly generated based on the Bernoulli distribution in an $n\times n$ matrix with  0 and 1 showing the presence and absence of a spot beam, respectively. By  vectorizing each encoded aperture as the row of a matrix, we get
\begin{equation}
\label{eq:e6}
\boldsymbol{\Phi}  =
\begin{bmatrix}
\Phi_{11} & \Phi_{12} & \ldots & \Phi_{1N} & \\
\Phi_{21} & \Phi_{22} & \ldots & \Phi_{2N}  & \\
\vdots & \vdots & \vdots & \vdots \\
\Phi_{M1} & \Phi_{M2} & \ldots & \Phi_{MN}
\end{bmatrix} \in  \textbf{C} ^{M \times N},
\end{equation}
where $\Phi_{ij}$, $1\leq i \leq M$ , $1\leq j\leq N$, are defined as
\begin{equation}
\label{eq:e7}
\Phi_{ij} =
\left\{
\begin{array}{ll}
0 \qquad  \text{$\textit{j}$th spot beam at the  $\textit{i}$th encoded aperture is absent} \\
1 \qquad  \text{ $ \textit{j}$th spot beam at the $\textit{i}$th encoded aperture is present}.
\end{array} \right.
\end{equation}\qquad
 In fact, this matrix shows whether the spot beams are active in both range and azimuth directions or not. Note that each encoded aperture normally forms an incomplete  image of a space object in each snapshot  and  we need $M = N$ encoded apertures to complete the image. However, because of the observation time constraint, we use $M < N$ encoded apertures and retrieve the images using the $L_{1}$ norm for space debris and the $TV$ norm for satellites.  As the space debris are physically smaller and their images are more sparse compared to those of a satellite, the use of $L_{1}$ norm is justified.  According to `\citep{Ros14}', $\boldsymbol {\Phi}$ should meet the RIP  condition as follows:
\begin{enumerate}
\item It is a non-square matrix with $M < N$,
\item Its entries are drawn from the Bernoulli distribution `\citep{Bar08}',
\item Its rows are orthogonal to have the least mutual coherence and further increase the probability of successful recovery `\citep{Can08}'. \end{enumerate} We represent the echo/signal received from active spot beams in each snapshot by the vector $\boldsymbol {y} \in C^{M}$. For example, $y_{1}$ is the echo signal in the first snapshot resulted from the first row of $\boldsymbol{\Phi}$, and in general, we can write
\begin{equation}
\label{eq:e8}
\boldsymbol{y}  = \boldsymbol{\Phi} \boldsymbol{x}.
\end{equation}
The above relationship defines an inverse problem which may be solved by the $L_{1}$ norm  for space debris. According to `\citep{Ros14}', smaller mutual coherence in the columns of $\boldsymbol {\Phi}$ leads to a more successful recovery of the unknown sparse vector $\boldsymbol {x}$ . Although, (8)  can be theoretically solved using the $L_{0}$ norm,  it is well-known that this problem is non-convex and thus NP-hard, and can be replaced by the $L_{1}$ norm given by `\citep{Zha06}'
\begin{equation}
\label{eq:e9}
\min \|\boldsymbol{x}\|_{1}  \qquad   s.t.   \quad    \boldsymbol{y} = \boldsymbol{\Phi}\boldsymbol{x},
\end{equation}
which is also known as the Basis Pursuit (BP) problem `\citep{Can06}'. On the other hand, to retrieve satellite images in orbit, we recommend the $TV$ norm defined  in isotropic form as the $L_{2}$ norm of the discrete gradient $ \nabla \boldsymbol{X} $ as `\citep{Jon20}'
\begin{equation}
\label{eq:e10}
\|\boldsymbol{X}\|_{TV} := \sum_{i,j} \| \nabla\boldsymbol{X}\|_{2}      \quad \nabla\boldsymbol{X}=
\begin{bmatrix}
D_{h} (\boldsymbol{X})
D_{v} (\boldsymbol{X})
\end{bmatrix},
\end{equation}
\noindent where  $D_{h} (\boldsymbol{X})$ and  $D_{v} (\boldsymbol{X})$ are the horizontal and vertical differences defined as
\begin{equation}
\label{eq:e11}
D_{h} (\boldsymbol{X}) =\boldsymbol{X}(i+1,j)-\boldsymbol{X}(i,j),
\end{equation}
\begin{equation}
\label{eq:e12}
D_{v} (\boldsymbol{X}) =\boldsymbol{X}(i,j+1)-\boldsymbol{X}(i,j).
\end{equation}
By this definition, the $TV$ norm incorporates the differences between the adjacent points of an image. In this work, we have used the NESTA algorithm `\citep{Jon20}' to solve the optimizations problems defined in (9) and (10). The flowchart of the method proposed for ISAR imaging of space debris and satellites is shown in Fig.\ref{fig:Fig.Flow}. In case of low resolution of the recovered image, the number of the encoded apertures increases and some steps are repeated. Note that when no space debris are detected, a satellite can be launched. Otherwise, the necessary measures should be taken from the Earth station.
\subsection{Observation time of space objects }	
The required time for observation of space objects includes spot beams generating and steering, data acquisition, and  processing time in each encoded aperture. This time totally takes less than $100 \mu$ and 10,000 spot beams  can be generated per second `\citep{Plo19}'. Thus, by considering   $ t_{s} $ as the time of each snapshot, the total time for $M$ snapshots is $ M t_{s} $, which should be less than the observation time for ISAR imaging, which is typically  $ 3 - 5$ seconds `\citep{Sci17}'. We will show that this value is  achievable by applying the proposed method.
\section{Simulation Results}	
The ISAR specifications shown in Table \ref{tab:Tab.1} are as close to the actual conditions as possible `\citep{Ang19}'.

\begin{table}
\tiny
  \centering
  \caption{
   ISAR specifications.}
  \label{tab:Tab.1}
\begin{tabular}{|l|l|}
\hline
Radar type & Pulse radar \\
\hline
Frequency band & X Band \\
\hline
Center frequency & 10.2 GHz\\
\hline
Bandwidth & 4.4 GHz \\
\hline
Transmitter signal & LFM \\
\hline
PRF & 200Hz \\
\hline
Pulse length &	$ 50$$ \mu s $\\
\hline
Encoded aperture dimensions  & 40 * 40\\
\hline
Number of possible spot beams, active or inactive, per snapshot &	$ 1600 $\\
\hline
Observation time &	3-5 seconds \\
\hline
Spot beams generating and processing time	& $ < 100 \mu s$ \\
\hline
Number of encoded apertures(M) &	$> 100 $  and $ < 1600 $ \\
\hline
Satellite orbital height &	$ 524 km \times 544 km$\\
\hline
Satellite speed &	$ 7430 m/s $ \\
\hline
Random distribution for $ \boldsymbol{\Phi} $  &	Bernoulli\\
\hline
\end{tabular}
\end{table}
\subsection{Effect of the number of  snapshots}	
\begin{figure}
\centering
\includegraphics[angle=0,width=0.8\textwidth]{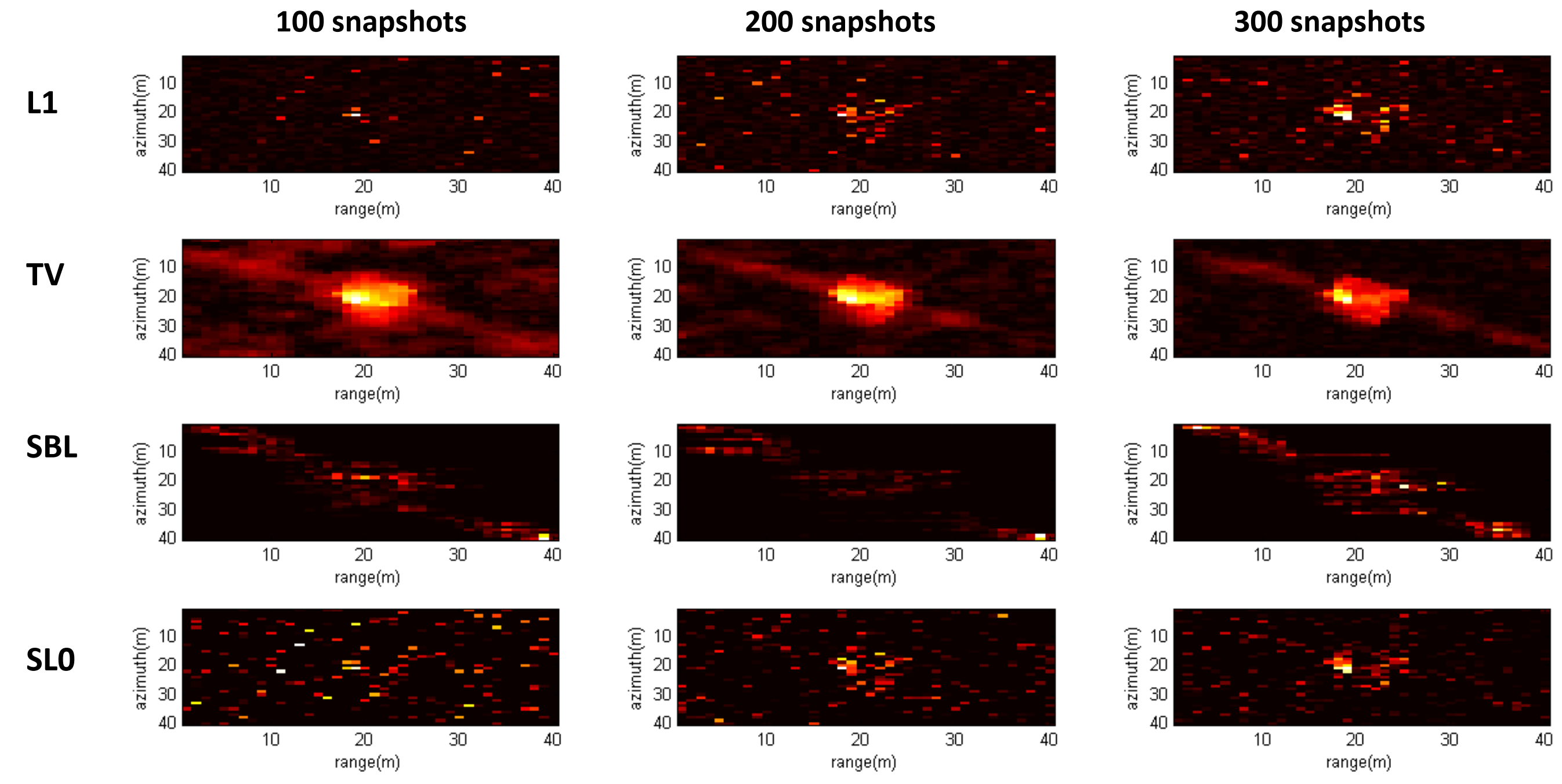}
\caption{Retrieved images for only satellite in orbit using $L_{1}$ ,$TV$, $SL_{0}$, and SBL for 100, 200, and 300 snapshots.}
\label{fig:Fig.snapshots}
\end{figure}

\begin{figure}[h]
\centering
\includegraphics[angle=0,width=0.8\textwidth]{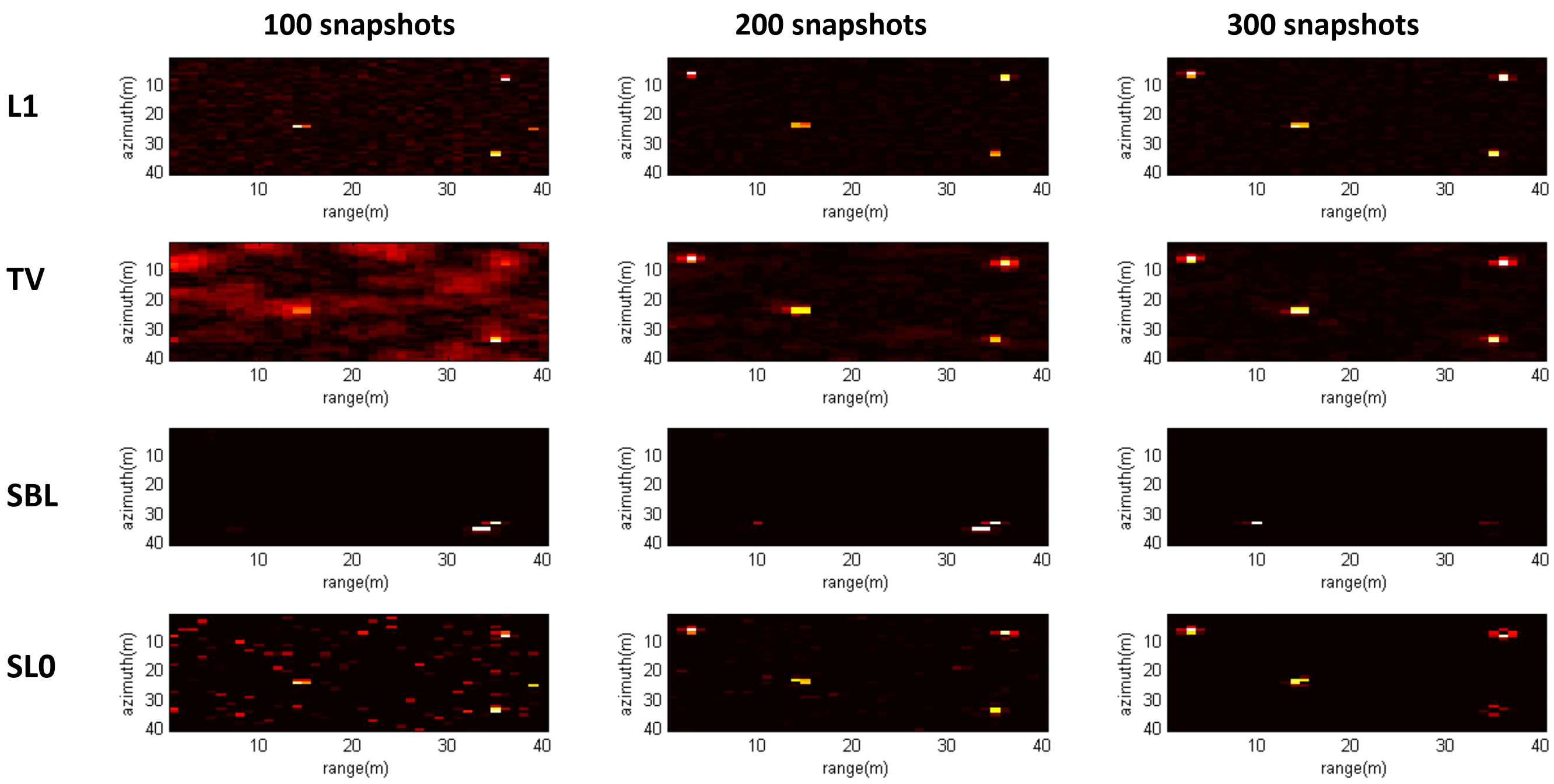}
\caption{Retrieved images for only space debris in  orbit using $L_{1}$ ,$TV$, $SL_{0}$, and SBL for 100, 200, and 300 snapshots.}
\label{fig:Fig.Before}
\end{figure}

We evaluate the performance of the proposed method for 100, 200, and 300 snapshots at SNR = 5 dB [5] in three different scenarios including; 1) only satellite in orbit, 2) only space debris in orbit, and 3) both satellite and space debris in orbit. Also, to recover the image, we consider $L_{1}$, $TV$, $SL_{0}$ norms, and SBL{Hui15}. Fig.\ref{fig:Fig.snapshots} shows the results for the first scenario. As seen, as opposed to the other methods, the $TV$ norm has retrieved acceptable images using a small number of snapshots. One can note that the SBL has also weakly recovered  the image, but it does impose a heavy computational burden compared to the $TV$ norm.
In the second scenario, we consider only space debris in orbit, for which the results are shown in Fig. \ref{fig:Fig.Before}. As seen, the $L_{1}$ norm has clearly recovered the space debris using only 100 snapshots which are lower than those of the other methods. This success can be justified by noting the fact that the images of space debris can be reasonably considered sparse.
In the third scenario, there exists a satellite in the vicinity of space debris in orbit. The results reveal that as opposed to the other methods, the $TV$ norm has acceptably recovered a satellite in the vicinity of space debris using a smaller number of snapshots.
\begin{figure}[h]
\centering
\includegraphics[angle=0,width=0.8\textwidth]{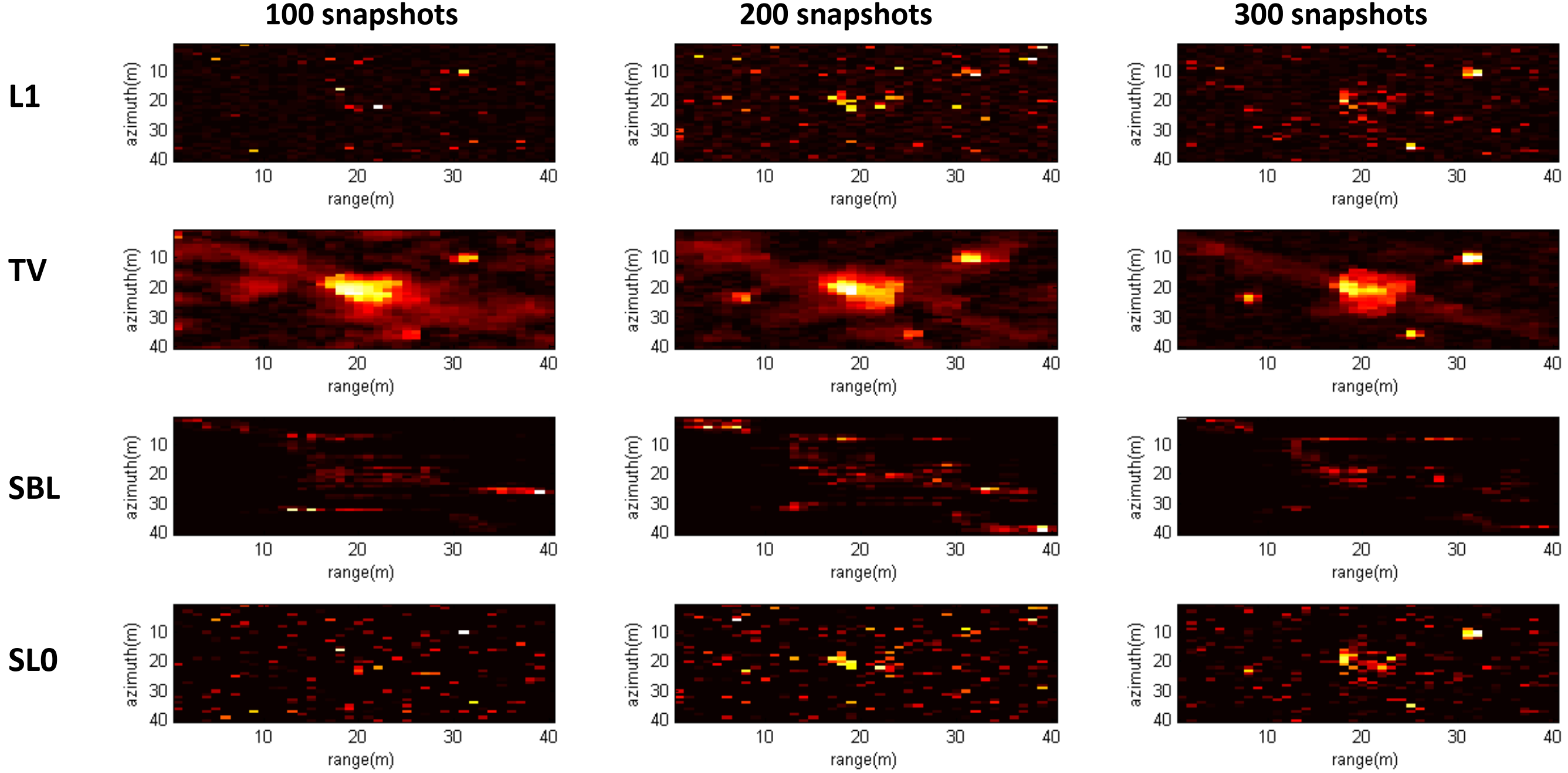}
\caption{Retrieved images for a satellite in the vicinity of space debris using $L_{1}$ ,$TV$, $SL_{0}$ norms, and SBL for 100, 200, and 300 snapshots.}
\label{fig:Fig.InOrbit}
\end{figure}
\subsection{Comparison of MSEs in image recovery}
\begin{figure}[h]
\centering
\includegraphics[angle=0,width=0.5\textwidth]{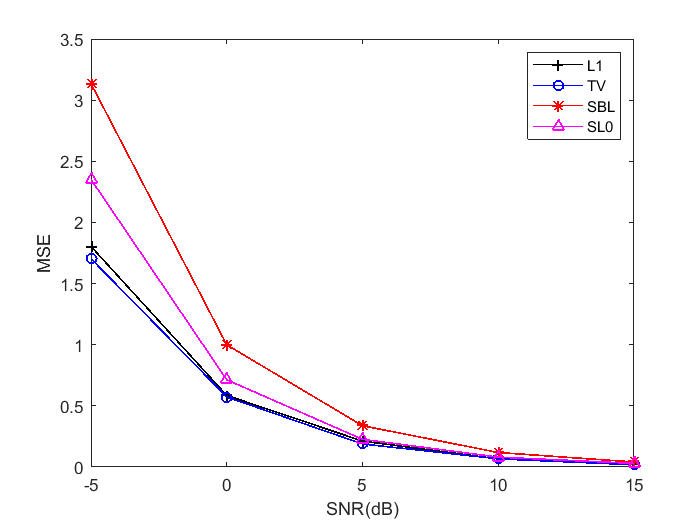}
\caption{MSEs of $L_{1}$, $TV$, $SL_{0}$ norms, and SBL for SNR= -5 to 15 dB. }
\label{fig:Fig.MSE}
\end{figure}

The Mean-Square Errors (MSEs) of different methods  are compared for only satellite in orbit at SNR = -5, 0, 5, 10, and 15 dB. The results are the average of 100 independent trials of the experiment. As seen in Fig.\ref{fig:Fig.MSE}, the MSEs for both $L_{1}$ and $TV$ norms are lower than those of the other methods at all SNRs.  These results are in agreement with the performance of the corresponding algorithms in Fig.\ref{fig:Fig.snapshots}.
\subsection{Performance of methods in presence of noise}
In the next simulation, both satellite and space debris are in orbit and image recovery is investigated in presence of  noise  at SNR= -5, 5, and 15 dB. It is observed that the $TV$ norm has acceptably recovered both objects at 5 dB while the other ones have failed, showing the better performance of the TV norm in such a scenario.
\begin{figure}[h]
\centering
\includegraphics[angle=0,width=0.8\textwidth]{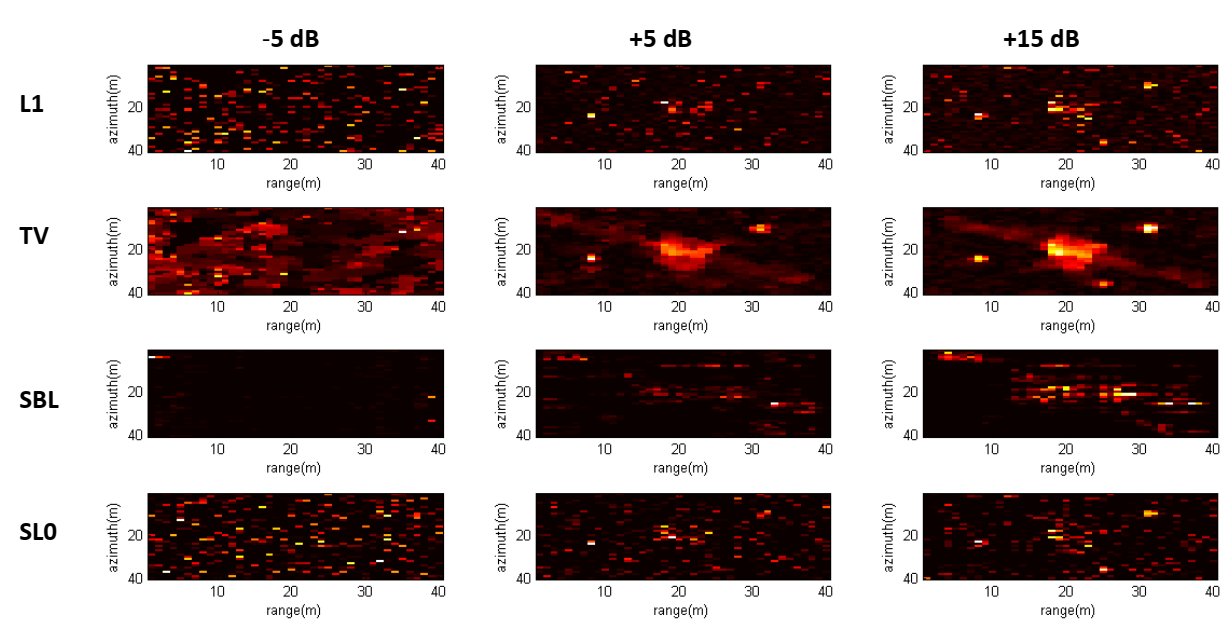}
\caption{ The Space debris retrieved images using four image recovery methods $L_{1}$, $TV$, and $SL_{0}$ norms and SBL at SNR = -5, +5, and +15 dB.}
\label{fig:Fig.snr}
\end{figure}
\subsection{Comparison of running times}
\begin{figure}[h]
\centering
\includegraphics[angle=0,width=0.5\textwidth]{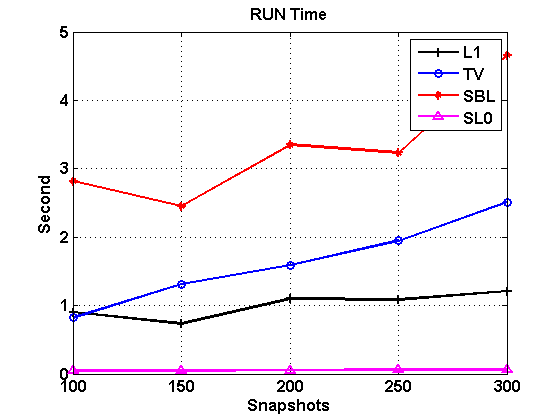}
\caption{ Running time of  $L_{1}$, $TV$, $SL_{0}$ norms and SBL for 100 to 300 snapshots.}
\label{fig:Fig.Elaps}
\end{figure}

The run times of the recovery methods are compared  in Fig.\ref{fig:Fig.Elaps} for 100 to 300 snapshots at 5 dB SNR. As shown, the SL0 is the fastest, then $L_{1}$ and $TV$ norms, and at last SBL has the slowest speed. However, it is imperative to note that although $SL_{0}$ is fast, its performance in image recovery is much lower for low snapshots and SNRs as already illustrated in Figs.\ref{fig:Fig.snapshots}-\ref{fig:Fig.MSE}. Also, SBL is very slow due to  incorporating the prior information into computations.
\section{Conclusion}
Given that space debris is a significant threat to satellites before launch or in orbit, we proposed  an encoded aperture method for ISAR imaging of space objects. Due to the fast rotation of space debris and having a limited time for observation, we focus on reducing the number of snapshots as much as possible. To do so, we used  a few spot beams in each snapshot  randomly  generated by the Bernoulli distribution. To recover the space object images, we applied $L_{1}$ and $TV$ norms. The performance of these methods was evaluated for a different number of snapshots and various SNRs.  Accordingly, the corresponding  MSEs and running times were compared. The simulation results showed that the  $TV$ norm can image a satellite in the absence/presence of space debris in orbit using a fewer number of snapshots compared to $L_{1}$, $SL_{0}$, and SBL methods. Also, the $TV$ norm performed better than the other methods at low SNRs by comparing the MSEs. Moreover, both $L_{1}$ and $TV$ norms were faster than the SBL in terms of running time. Of course, $SL_{0}$ achieved much faster speed than the others, but at the cost of generating greatly low-resolution images that are mostly unusable. On the other hand, $L_{1}$ outperformed the SBL for space debris in the absence of  satellites due to the sparse identity of the respective images. These results led us to design an ISAR imaging procedure by using the $L_{1}$ norm for space debris with no satellite in orbit and the $TV$ norm for satellites in the absence/presence of space debris in orbit.

\end{document}